\documentclass[preprint2]{aastex}

\usepackage{epsfig}
\usepackage{graphics}

\tighten

\shorttitle{Resolved CO Emission at z=6.42}
\shortauthors{Walter et al.}

\begin{document}

\title{Resolved Molecular Gas in a Quasar Host Galaxy at Redshift z=6.42}

\author{Fabian Walter\altaffilmark{1,2}, Chris Carilli}
\affil{National Radio Astronomy Observatory, P.O. Box O, Socorro, NM
87801, USA}

\author{Frank Bertoldi, Karl Menten}
\affil{Max Planck Institut f\"ur Radioastronomie, Auf dem H\"ugel 69;
53121 Bonn, Germany}

\author{Pierre Cox}
\affil{Institut d'Astrophysique Spatiale, Universite de Paris-Sud, 91405
Orsay, France}

\author{K.Y. Lo}
\affil{National Radio Astronomy Observatory, 520 Edgemont Road
Charlottesville, VA 22903, USA}

\author{Xiaohui Fan}
\affil{Steward Observatory, University of Arizona, 933 N. Cherry Ave.,
Tucson, AZ 85721, USA}

\author{Michael A. Strauss}
\affil{Princeton University Observatory, Princeton, NJ 08544, USA}

\altaffiltext{1}{Jansky Postdoctoral Fellow at the National Radio 
Astronomy Observatory}
\altaffiltext{2}{Current address: Max Planck Institute for Astronomy, K\"onigstuhl 17, 69117 Heidelberg, Germany}

\begin{abstract}
  
  We present high-resolution VLA observations of the molecular gas in
  the host galaxy of the highest redshift quasar currently known,
  SDSS\,J1148+5251 (z=6.42). Our VLA data of the CO(3--2) emission
  have a maximum resolution of 0.17$''\times0.13''$ ($\leq$1\,kpc),
  and enable us to resolve the molecular gas emission both spatially
  and in velocity. The molecular gas in J1148+5251 is extended to a
  radius of 2.5\,kpc, and the central region shows 2 peaks, separated
  by 0.3$''$ (1.7\,kpc). These peaks account for about half of the
  total emission, while the remainder is more extended. Each of these
  unresolved peaks contains a molecular gas mass of
  $\sim5\times10^9$\,M$_{\odot}$ (similar to the total mass found in
  nearby ULIRGS) and has an intrinsic brightness temperature of
  $\sim35$\,K (averaged over the 1\,kpc-sized beam), comparable to
  what is found in nearby starburst centers.  Assuming that the
  molecular gas is gravitationally bound, we estimate a dynamical mass
  of $\sim4.5\times10^{10}$\,M$_{\odot}$ within a radius of 2.5\,kpc
  ($\sim5.5\times10^{10}$\,M$_{\odot}$ if corrected for a derived
  inclination of $i\sim65^{\circ}$). This dynamical mass estimate
  leaves little room for matter other than the detected molecular gas,
  and in particular the data are inconsistent with a
  $\sim10^{12}$\,M$_\sun$ stellar bulge which would be predicted based
  on the M$_{\rm BH}-\sigma_{\rm bulge}$ relation. This finding may
  indicate that black holes form prior to the assembly of the stellar
  bulges and that the dark matter halos are less massive than
  predicted based on the black hole/bulge mass relationship.
\end{abstract}


\section{Introduction}
\label{par:intro}

More and more objects have been found at the highest redshifts in
recent years, back to the 'Dark Ages' (the epoch of formation of the
first luminous structures), at z$>$6 when the universe was less than a
gigayear old (e.g., Fan et al.\ 2002, 2003, 2004, Hu et al.\ 2002,
Stanway et al.\ 2003). Optical spectra of the brightest quasars at
z$>$6 show a clear Gunn \& Peterson (1965) effect, demonstrating that
these objects are located at the end of cosmic reionization (Becker et
al.\ 2001, Fan et al.\ 2003, White et al.\ 2003). It is of paramount
importance to study these sources in detail to measure the stellar and
gaseous constituents, chemical abundances and dynamical masses of the
host galaxies, which in turn constrain galaxy evolution models in the
very early universe. Studies of the host galaxies of bright quasars
are extremely challenging in the optical and/or near-infrared as the
central active galactic nucleus (AGN) greatly overshines its
surrounding.  Radio observations of the molecular gas phase, however,
are a promising method to measure such important properties of distant
quasar host galaxies as the total quantity of molecular gas and the
dynamical masses.

The highest redshift quasar currently known is
SDSS\,J114816.64+525150.3 (hereafter: J1148+5251) at z=6.42 (Fan et
al.\ 2003). J1148+5251 is a highly luminous object which is thought to
be powered by mass accretion onto a supermassive black hole of mass
1--5$\times10^9$ M$_{\odot}$ (which accrets at the Eddington limit,
Willot et al.\ 2003).  Thermal emission from warm dust has been
detected at millimetre wavelengths, implying a far infrared (FIR)
luminosity of $1.3\times 10^{13}$\,L$_\odot$ (Bertoldi et al.\ 2003a),
corresponding to about 10$\%$ of the bolometric luminosity of the
system. J1148+5251 falls on the radio-FIR correlation found for nearby
starburst galaxies (Carilli et al.\ 2004); if the heating of the dust
is dominated by young stars, then the implied star formation rate is
an impressive $\sim3000$\,M$_{\odot}$\,yr$^{-1}$. High--resolution HST
imaging does not reveal multiple imaging of the central AGN (White et
al.\ 2004), rendering strong magnification unlikely (cf. the multiple
optical imaging in the gravitationally lensed system PSS2322+1944,
Carilli et al.\ 2003).

Using the Very Large Array (VLA) and the Plateau de Bure
interferometer (PdBI), we have detected various rotational lines of
carbon monoxide (CO), the most common tracer of molecular gas, in
J1148+5251 (Walter et al. 2003, Bertoldi et al. 2003b). This discovery
demonstrated that abundant, dense and warm molecular gas (the
requisite fuel for star formation) is already present at these large
lookback times (a redshift of z=6.4 corresponds to an age of the
universe of 870\,Myr in a cosmology with
H$_0\!=\!71$\,km\,s$^{-1}\,$Mpc$^{-1}$, $\Omega_{\Lambda}$=0.73,
$\Omega_{m}$=0.27 which we adopt in the following).

Here we present follow-up high-resolution observations of the
molecular gas distribution of J1148+5251 obtained with the NRAO
VLA\footnote{The National Radio Astronomy Observatory is a facility of
the National Science Foundation operated under cooperative agreement
by Associated Universities, Inc.}. These observations resolve, for the
first time, the host galaxy of this luminous quasar on kiloparsec
scales (1\,kpc = 0.18$''$ at z=6.42) and enable us to constrain its
molecular gas properties and the dynamical mass.

\section{Observations}

Using the VLA in D configuration, Walter et al.\ (2003) detected the
CO(3--2) line ($\nu_{\rm rest}$=\,345.796\,GHz) in J1148+5251, which
is shifted to 46.61\,GHz (6.43\,mm) at z=6.419. The emission was
unresolved in the D-array observations at 1.5$''$ resolution but the
line strength of 0.6\,mJy indicated that follow-up VLA observations at
higher spatial and spectral resolution were possible. In addition,
surface brightness arguments yielded a minimum size of 0.2$''$(T$_{\rm
B}$[50K])$^{-0.5}$, a resolution which can be reached with the VLA
B--array.  We therefore performed follow-up observations of J1148+5251
at the VLA in B array (60 hours) and C array (24 hours) from January
-- April 2004 (total time on source: $\sim$60 hours). Observations
were done in fast switching mode using the nearby source 11534+49311
for secondary amplitude and phase calibration; the phase stability in
all runs was excellent ($\sim10^{\circ}$ phase rms on 10\,km
baselines).

Two 25\,MHz intermediate frequencies (IFs, centered at 46.603250\,GHz
and 46.622050\,GHz, respectively) with 7 channels each (channel width:
3.1250\,MHz) were observed simultaneously, leading to a total coverage
of 37.5 MHz (or 240\,km\,s$^{-1}$, excluding the overlap and edge
channel). This velocity range encompasses most of the CO line-width
measured by the VLA and the PdBI (FWHM=280\,km\,s$^{-1}$), but does
not include the line wings and the continuum (in earlier observations
we derived a $2\sigma$ upper limit of the continuum at 46\,GHz of
0.1\,mJy, Walter et al.\ 2003). Given the limited signal to noise
ratio in individual channels, three individual channels were binned to
one channel, resulting in a total of 4 independent channels
(width=9.375\,MHz, or 60\,km\,s$^{-1}$).

For our analysis we created two datasets. The first dataset is based
on the new C and B array data with a UV taper of 1\,M$\lambda$,
resulting in a beamsize of $0.35''\times0.30''$ and an rms noise of
86$\mu$Jy\,beam$^{-1}$ in a 9.375\,MHz channel. The rms noise in the
map covering the entire bandpass (37.5\,MHz) is
43$\mu$Jy\,beam$^{-1}$. To boost the resolution in the central region,
the B array data only were mapped using 'Natural' weighting; this
results in a beam of $0.17''\times0.13''$ and an rms noise of
45$\mu$Jy\,beam$^{-1}$. The original D array data were not included in
this analysis as they were not observed in spectral line mode.


\vspace{1cm}

\section{Results}

\subsection{Global Distribution of Molecular Gas}

In Fig.~1 we present the CO(3--2) emission in J1148+5251 over the
total measured bandwidth (37.5\,MHz, 240\,km\,s$^{-1}$) at
$0.35''\times0.30''$ (1.9\,kpc$\times$1.7\,kpc) resolution. The
emission is clearly extended and Gaussian fitting in the map plane
gives a deconvolved major axis (FWHM) of 0.65$\pm0.12''$
($\sim$3.6\,kpc), a marginally resolved minor axis of 0.25$\pm0.12''$
($\sim$1.4\,kpc) and a position angle of 15$\pm10^\circ$ (measured
east from north). Molecular gas can be seen out to distances of
0.42$''$ from the centre ($\sim$2.5\,kpc, numbers are deconvolved for
the beam size).  The source is possibly extended towards the north
(see also Fig.~2, first channel map) but more sensitive observations
are needed to confirm this extension. If we assume that the main
molecular gas concentration forms an inclined disk, this gives an
inclination of $\sim$65$^\circ$ (where 0$^\circ$ corresponds to
face--on) which implies that any inclination correction to the
measured rotational velocities is minimal. The fitted peak to the
central distribution is 0.21 mJy\,beam$^{-1}$ and the integral
intensity is 0.55\,mJy. This corresponds to an H$_2$ mass of
M(H$_2)\sim1.6\times10^{10}$\,M$_\odot$.\footnote{Here and in the
following we assume constant brightness in the CO(3--2) and CO(1--0)
transitions and a conversion factor to convert CO luminosities to
H$_2$ masses of $\alpha \sim 0.8$
M$_\odot$\,(K\,km\,s$^{-1}$\,pc$^{2})^{-1}$, see Walter et al. 2003.}
A 1\,mJy continuum point source 40$''$ north-east of the QSO is
unresolved, emphasizing that phase decoherence is not the reason that
the emission of J1148+5251 appears extended.

Given our beamsize of 0.35$''\times$0.30$''$, the peak brightness of
0.21 mJy\,beam$^{-1}$ corresponds to a surface brightness of 1.1\,K,
or a beam--smoothed brightness temperature of
1.1$\times(1+z)$=8.3\,K at z=6.42.

\subsection{Dynamics of the Molecular Gas}

Assuming that the molecular gas is gravitationally bound and forms an
inclined disk with a radius of 2.5\,kpc (see previous section) and
that the gas seen in the PdBI spectrum emerges from the same region
(full width at zero intensity: 560\,km\,s$^{-1}$ or v$_{\rm
rot}\sim$280\,km\,s$^{-1}$, Bertoldi et al.\ 2003b), we derive an
approximate dynamical mass for J1148+5251 of
$\sim4.5\times10^{10}$\,M$_\odot$ (or
$\sim5.5\times10^{10}$\,M$_\odot$ if we correct for an inclination of
$\sim$65$^\circ$) with an error of order 50\%. Within the large
uncertainties, this number is compatible with the derived molecular
gas mass and the mass of the central black hole, but does not leave
much room for additional matter (e.g. a massive stellar bulge, see
discussion below).

Using the velocity information in our new observations, we can now
investigate the spatially resolved dynamics in the centre of
J1148+5251. Fig.~2 shows four channel maps (width of one channel:
9.375\,MHz, or 60\,km\,s$^{-1}$) which cover the entire observed
bandpass at $0.35''\times0.30''$ (2.0\,kpc$\times1.7$\,kpc)
resolution. Due to the intrinsic faintness of the emission, the
channel maps are noisy, however some interesting features can still be
identified. In particular there is some indication that emission is
moving from the north to the south with increasing channel number
(i.e., increasing frequency). Observations with better sensitivity
(and covering more of the total CO line--width) are required to
further constrain the internal dynamics of J1148+5251.

\subsection{Distribution of CO at 1\,kpc scale}

The distribution of molecular gas at even higher resolution reveals
that the situation is likely more complex than described by the simple
disk model above. In Fig.~3 we show our highest resolution CO(3--2)
map derived from B-array data only. Here the resolution is only
$0.17''\times0.13''$, or $\sim$0.95\,kpc$\times0.72$\,kpc.\footnote{We
note that this is comparable to resolutions typically achieved with
single-dish telescopes when studying galaxies in the nearby universe
in the CO(1--0) transition.} This map shows that the emission breaks
up into two regions, a northern source at 11$^{\rm h}$48$^{\rm
m}$16.640, 52$^\circ$51$'$50.51 (peak flux density:
192$\pm45$\,$\mu$Jy), and a southern source at 11$^{\rm h}$48$^{\rm
m}$16.641, 52$^\circ$51$'$50.21 (flux density: 180$\pm45$\,$\mu$Jy),
i.e., separated by $\sim$0.3$''$ (1.7\,kpc).  On a Kelvin scale, both
sources are at $\sim$4.5\,K, which corresponds to a lower limit for
the beam--averaged brightness temperature of $\sim$35\,K at z=6.42.

To interpret this morphology and its relation to the optical quasar,
astrometric precision to better than 0.05$''$ is needed. However, at
the scales under consideration it proves to be extremely difficult to
tie the radio and optical frames together. This situation is further
complicated by the fact that J1148+5251 has a very atypical spectrum
(showing a complete Gunn Peterson effect in the optical and only
covering parts of the Sloan z band). Thus differential chromatic
aberration would preclude tying the SDSS z-band image of J1148+5251 to
the radio frame to better than a few tenths of an arcsecond even if
common sources existed in both frames. This uncertainty is indicated
by the size of the cross in all figures. As the southern CO emission
is situated more towards the centre of the entire molecular gas
distribution (cf. Fig.~1) we consider it likely, for symmetry reasons,
that the optical QSO is associated with the southern source.  In this
picture the northern source may possibly be the nucleus of an object
in the process of merging with the quasar host. This situation may be
reminiscent of the molecular gas distribution around the z=4.7 QSO
BRI\,1202-0725 in which two molecular gas peaks have been detected as
well (albeit on a factor 10 larger scale, Omont et al.\ 1996, Carilli
et al.\ 2002). Alternatively, the northern source could be a region of
high surface brightness within J1148+5251.

\section{Summary and Discussion}

We present the first resolved maps of a system located at the end of
cosmic reionization at z=6.4. Our high-resolution VLA data enable us
to resolve the molecular gas emission in the host galaxy both
spatially and in velocity space.

The molecular gas distribution in J1148+5251 is extended out to radii
of 2.5\,kpc. The central region is resolved and shows 2 peaks,
separated by 1.7\,kpc; they account for about half of the total
emission, with the other half present in the more extended molecular
gas distribution. Each of these peaks harbors a molecular gas mass of
$\sim5\times10^9$\,M$_{\odot}$ within a radius of 0.5\,kpc,
respectively; this mass is similar to the total mass found in nearby
ULIRGS such as Mrk\,273 or Arp\,220 (Downes \& Solomon 1998). The
peaks have intrinsic brightnesses of $\sim35$\,K (averaged over the
1\,kpc-sized beam), similar to what is found in the centres of nearby
active galaxies (Downes \& Solomon 1998, assuming constant surface
brightness down to the CO(1--0) transition), albeit measured over a
larger physical area.

We have assumed that the gas is gravitationally bound and is situated
in a disk, although the data do not rule out the presence of an
unbound system, such as an ongoing merger. Based on the extent of the
molecular gas distribution and the line-width measured from the higher
CO transitions we derive a dynamical mass of
$\sim4.5\times10^{10}$\,M$_{\odot}$
($\sim5.5\times10^{10}$\,M$_{\odot}$ if we correct for an inclination
of $i\sim65^{\circ}$).  This dynamical mass estimate can account for
the detected molecular gas mass within this radius but leaves little
room for other matter. In particular, given a black hole mass of mass
$\sim$1-5$\times10^9$ M$_{\odot}$ (Willot et al.\ 2003), this
dynamical mass could not accomodate an order few$\,\times
10^{12}$\,M$_{\odot}$ stellar bulge which is predicted by the
present--day M$_{\rm BH}-\sigma_{\rm bulge}$ relation (Ferrarese \&
Merritt 2000, Gebhardt 2000), if this relation were to hold at these
high redshifts. Even if we assume a scenario in which this bulge was
10 times the scale length of the molecular gas emission, we would
still expect a bulge contribution of few$\,\times
10^{11}$\,M$_{\odot}$ within the central 2.5\,kpc (assuming a central
density--profile of $\rho\sim r^{-2}$, e.g., Jaffe 1983, Tremaine
1994) which can not be reconciled with our results. Our finding
therefore suggests that black holes may assemble before the stellar
bulges. This would be in conflict with the popular picture of co-eval
evolution of the central black hole and the stellar bulge. The derived
dynamical mass and the hypothetical stellar bulge mass (based on
M$_{\rm BH}-\sigma_{\rm bulge}$) can be only brought in agreement if
(i) our derived inclination is significantly off (i.e., $i<10^\circ$),
(ii) the estimated black hole mass is over-estimated by orders of
magnitudes (which would imply highly super-Eddington accretion), or
(iii) if the assumption that the gas is gravitationally bound is
seriously wrong. We note, however, that depending on the space density
of similar sources at these high redshifts, the smaller dynamical
masses suggested in this study may be in better agreement with the
masses predicted by CDM simulations in the very early universe as a
10$^{12}$\,M$_{\odot}$ bulge would imply a rather massive dark matter
halo of $>10^{14}$\,M$_{\odot}$.

This study shows the potential for future studies of the molecular gas
content and dynamical masses in the highest redshift galaxies using
ALMA, where resolutions of $<0.1''$ will be achieved routinely.

\acknowledgements{
We would like to thank the anonymous referee for very helpful
comments.  }

\clearpage

\begin{figure}[t!]
\epsscale{1}
\plotone{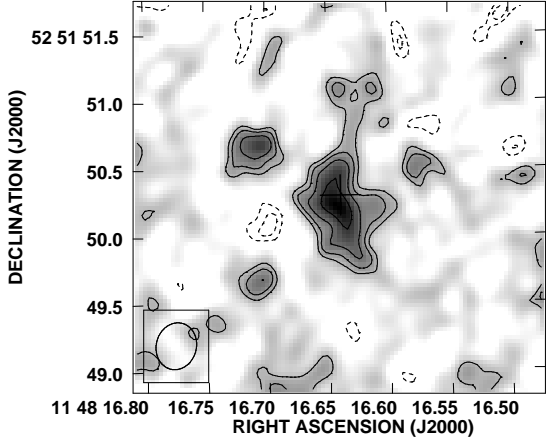}
\caption{CO(3--2) map of J1148+5251 of the combined B-- and C--array
dataset (covering the total bandwidth, 37.5\,MHz or
240\,km\,s$^{-1}$). Contours are shown at -2, -1.4, 1.4, 2, 2.8,
4$\times \sigma$ (1$\sigma$=43$\mu$Jy\,beam$^{-1}$). The beamsize
($0.35\times0.30''$) is shown in the bottom left corner; the cross
indicates the SDSS position (and positional accuracy) of J1148+5251.}
\end{figure}

\begin{figure}[b!]
\epsscale{2}
\plotone{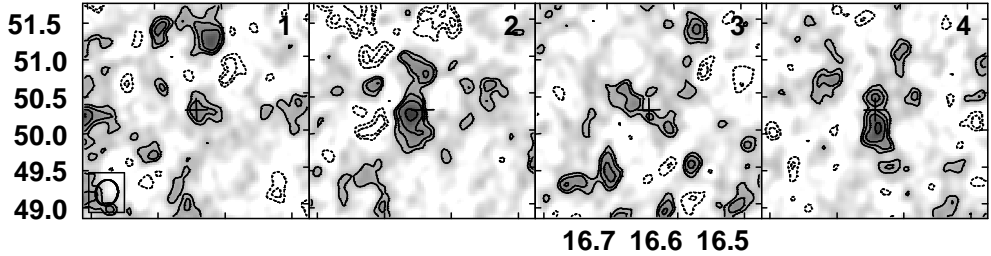}
\caption{Channel maps of the CO(3--2) emission based on the combined B
  and C array data sets. The same region is shown as in Fig.~1
  (Declination is in arcsecond relative to 52$^{\circ}$ 51$'$ and the
  Right Ascension, labeled on the third channel, is relative to
  11h~48m). One channel width is 9.375\,MHz, or 60\,km\,s$^{-1}$
  (frequencies increase with channel number and are shown at
  46.597000, 46.606375, 46.615750 and 46.625125\,GHz).  Contours are
  shown at -2.8,s -2, -1.4, 1.4, 2, 2.8, 4$\times \sigma$
  (1$\sigma$=86$\mu$Jy\,beam$^{-1}$). The beamsize
  ($0.35\times0.30''$) is shown in the bottom left corner; the cross
  indicates the SDSS position (and positional accuracy) of
  J1148+5251.}

\end{figure}

\clearpage

\begin{figure}[t!]
\epsscale{1}
\plotone{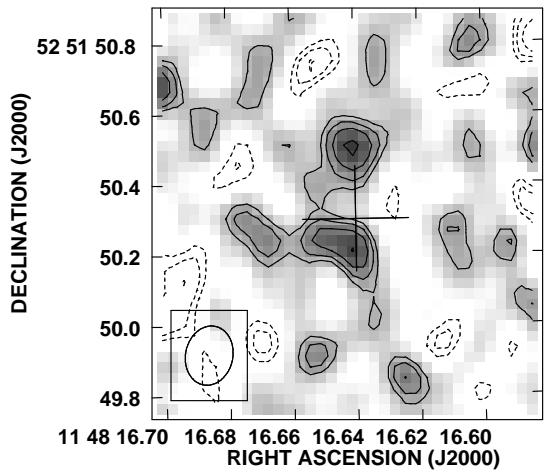}
\caption{CO(3--2) map at $\sim$1\,kpc resolution of 
the central region displayed in Figs.~1 and 2.  Contours are shown at
-2.8, -2, -1.4, 1.4, 2, 2.8, 4$\times \sigma$
($\sigma$=45$\mu$Jy\,beam$^{-1}$). The beamsize ($0.17\times0.13''$)
is shown in the bottom left corner. The positional uncertainty of the
SDSS position is of order $\pm 0.1''$ or about the size of the cross.}

\end{figure}

\end{document}